# A Hough Transform based Technique for Text Segmentation

Satadal Saha, Subhadip Basu, Mita Nasipuri and Dipak Kr. Basu

**Abstract**—Text segmentation is an inherent part of an OCR system irrespective of the domain of application of it. The OCR system contains a segmentation module where the text lines, words and ultimately the characters must be segmented properly for its successful recognition. The present work implements a Hough transform based technique for line and word segmentation from digitized images. The proposed technique is applied not only on the document image dataset but also on dataset for business card reader system and license plate recognition system. For standardization of the performance of the system the technique is also applied on public domain dataset published in the website by CMATER, Jadavpur University. The document images consist of multi-script printed and hand written text lines with variety in script and line spacing in single document image. The technique performs quite satisfactorily when applied on mobile camera captured business card images with low resolution. The usefulness of the technique is verified by applying it in a commercial project for localization of license plate of vehicles from surveillance camera images by the process of segmentation itself. The accuracy of the technique for word segmentation, as verified experimentally, is 85.7% for document images, 94.6% for business card images and 88% for surveillance camera images.

**Index Terms**—Binarization, Connected Component Labeling, Hough Transform, Segmentation

———————— ◆ ————————

## 1 INTRODUCTION

SEGMENTATION of text line from images is a major and critical component of an Optical Character Recognition (OCR) system. An OCR system takes an image as input and generates a character set in editable form as an output. The image is first peprocessed and then it is passed through the process of binarization, line segmentation, word segmentation, character segmentation and character recognition. Text segmentation, in general, incorporates line segmentation, word segmentation and character segmentation from a document image. It is the process through which the text component within an image is isolated from the background. For proper reconstruction of the editable text lines from the recognized characters, the line of text is first segmented, then from the segmented line the words are segmented and then from that the characters are segmented.

Many research works have already been done through last two and half decades regarding different aspects of OCR system. The research initially started with printed text segmentation and recognition and later the works have been extended to segmentation and recognition ofhandwritten texts also. The main application of OCR was initially to recognize scanned document images. But with the progress of automation, different technical areas have been evolved, which require expertised OCR system. The area may be a business card reader (BCR) system where the image of a business card is acquired through even a mobile camera and then the texts therein are segmented, recognized and stored in a database for easy access. It may also be a vehicle license plate recognition (LPR) system where from the image of a general scene containing vehicles, the license plate is located, characters are segmented and then they are recognized for maintenance of traffic management system.

The past research works were mainly focused on scanned document images. Namboodiri etal [1] reported a work on online handwritten character recognition of six types of scripts. The method uses vertical projection of the lines and uses the inter-line distance for line segmentation. The words are segmented from the lines using the horizontal projection of the lines and uses the inter-word distance. But the technique may not be suitable for considerably skewed lines. Text line extraction from multi-skewed handwritten document images has been reported by S. Basu etal [2]. The work implements a water flow technique for extraction of text lines with high rate of success. A. Khandelwal etal [3] reported a work on text line segmentation from unconstrained handwritten document images by comparing neighborhood connected components. The method is applied over four types of scripts with high rate of success for text line extraction.

The techniques reported earlier provides high rate of accuracy in case of text segmentation from document images but they may not always suitable for camera captured images because of poor resolution and non-uniform illumination of the image [4]. The skewness of the projection plane of the camera and the front plane of the object causes distortion of texts in camera captured images. In the proposed approach the classical Hough transform is

————————————————
- S. Saha is with the Dept. of Computer Science and Engg., Mckv Institute of Engg., Howrah
- S. Basu is with the Dept. of Computer Science and Engg., Jadavpur University, Kolkata
- M. Nasipuri is with the Dept. of Computer Science and Engg., Jadavpur University, Kolkata
- D. K. Basu is with the Dept. of Computer Science and Engg., Jadavpur University, Kolkata

used tactfully for directional segmentation of lines and words from any type of images. The Hough image is generated from the binarized edge map of the image and the parameters of the Hough transform are tuned depending on the variety of the images used in BCR or LPR or general document image recognition system.

## 2 ALGORITHM AND NOTATION

### 2.1 Hough Transform

In automated analysis of digital images, a frequemtly arising problem is detecting the simple shapes like straight line, circle or ellipse. In most of the cases an edge detector can be used as a pre-processing stage to obtain image points or image pixels that are on the desired curve in the image space. But due to imperfections in either the image data or the edge detector there may be missing or isolated or disjoint points or pixels on the desired curves as well as there may be spatial deviations between the ideal line or circle or ellipse and the noisy edge points as obtained from the edge detector. For these reasons, it is often non-trivial to group the extracted edge features to an appropriate set of lines, circles or ellipses. The purpose of the Hough transform is to address this type of problem by making it possible to perform groupings of edge points into object candidates by performing an explicit voting procedure over a set of parameterized image objects. The Hough transform is briefly described below.

Let us consider a single isolated edge point $(x, y)$ in the image plane. There could be an infinite number of lines that could pass through this point. Each of these lines can be characterized as the solution to some particular equation. In the simplest form a line can be expressed in the slope-intercept form as

$$y = mx + c$$

where, $m$ is the slope of the line w.r.t. x axis and $c$ is the intercept on y axis made by the line. Any line can be characterized by these two parameters pair $(m, c)$. For all the lines that pass through a given point $(x, y)$, there is a unique value of $c$ for $m$, given by

$$c = y - mx$$

The set of $(m, c)$ values corresponding to the lines passing through point $(x, y)$ form a line in $(m, c)$ space. Every point in image space $(x, y)$ corresponds to a line in parameter space $(m, c)$ and in the reverse way, each point in $(m, c)$ space corresponds to a line in image space $(x, y)$.

The Hough transform works by letting each feature point $(x, y)$ vote in $(m, c)$ space for each possible line passing through it. These votes are totalled in an accumulator.

Suppose that a particular $(m, c)$ has one vote — this means that there is a feature point through which this line passes. If it has two votes then it means that two feature points lie on that line. If a position $(m, c)$ in the accumulator has $n$ votes, this means that $n$ feature points lie on that line.

### The Hough Transform Algorithm

1. Find all of the desired feature points in the image space.
2. For each feature point in image space
3. For each possibility $i$ in the accumulator that passes through the feature point
4. Increment that position in the accumulator
5. Find local maxima in the accumulator.
6. On requirement map each maxima in the accumulator back to image space.

### Alternative representation of Lines

The slope-intercept form of a line as discussed above-has a problem with vertical lines: both $m$ and $c$ are infinite. To eliminate this problem of representing the point in the $(m, c)$ space another way of expressing a line is in $(\rho, \theta)$ form is used as

$$x \cos \theta + y \sin \theta = \rho$$

One way of interpreting this is to drop a perpendicular from the origin to the line. $\theta$ is the angle that the perpendicular makes with the *x*-axis and $\rho$ is the length of the perpendicular. $\theta$ is bounded by $[0, 2\pi]$ and $\rho$ is bounded by the diagonal of the image. Instead of making lines in the accumulator, each feature point votes for a sinusoid of points in the accumulator. Where these sinusoids cross, there are higher accumulator values. Finding maxima in the accumulator still equates to finding the lines.

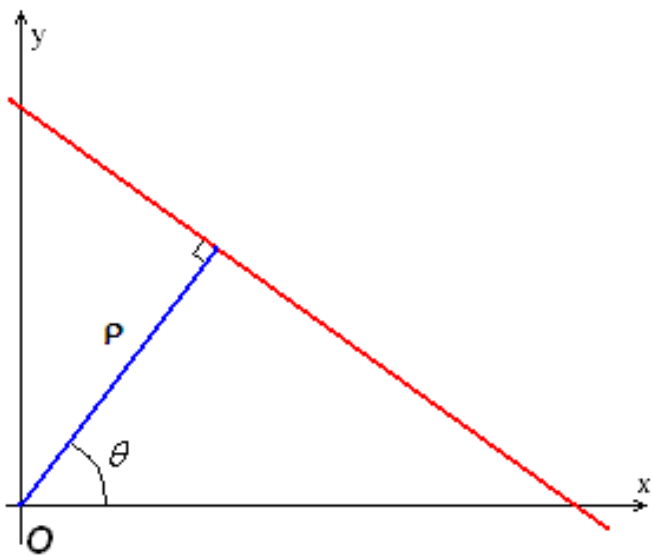

Fig. 1. Alternative representation of straight line in (ρ, θ) format.

The $(\rho, \theta)$ plane is sometimes referred to as *Hough space* for the set of straight lines in a two dimensional image space. The steps of implementation can be summarized below.

- For each image data point, a number of lines are plotted going through it, all at different angles.
- For each line a line is plotted which is perpendicular to it and which intersects the origin.
- The length and angle of each dashed line is measured.
- These the steps are repeated for each data point.
- A graph of length against angle, known as a Hough space graph, is then created.

For the line-matching Hough Transform, the orientation of the line is one of the parameters. If the orientation parameter is not used then matches in a specific orientation can be found out. The orientation parameter can be changed sequentially in an incremental way to find out all the lines oriented in different directions.

### 2.2 Work Flow Diagram

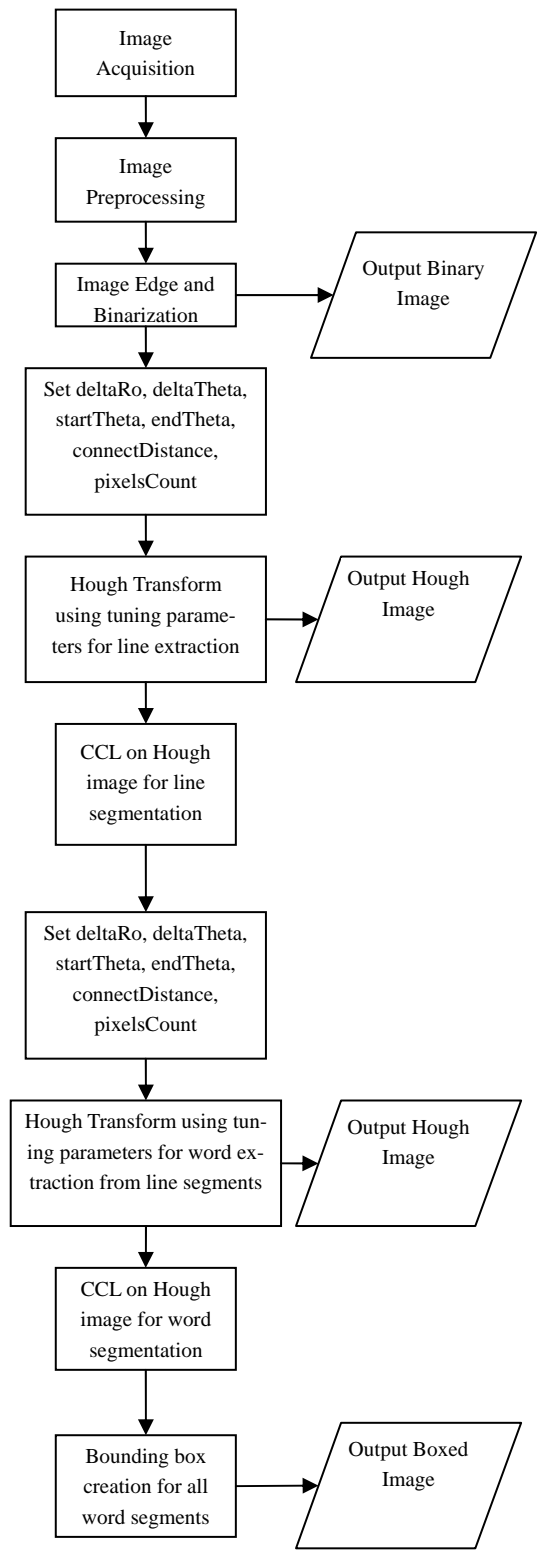

Fig. 2. Schematic work flow of the proposed work

Image acquisition is done through scanner or mobile camea or CCTV camera. These images are preprocessed to remove noises. The preprocessed images are then converted to grey scale image and subsequently binarized to form black and white image containing white background and black foreground. Hough transform with tuned parameter set is then applied on the binarized image to get the text lines as connected components in the Hough image. All the text lines are then segmented using connected component labeling (CCL) algorithm. Once text line segmentation is done, Hough transform is again applied with differently tuned parameter set to get the words as individual segments. The words are then segmented using CCL algorithm. This flow of work as illustrated in Fig. 2. is applied on three different domains of applications: BCR, LPR and document image recognition with minute changes in implemnation.

### 2.3 Image Acquisition

In case of BCR, the business card is placed in front of a hand held mobile phone and the color image is captured using the camera of the mobile phone. The image is then transferred to the PC using either cable or through bluetooth device. In case of LPR, camera is placed at road side and color images are captured at regular interval of time. The images contain front face of vehicles standing before the stop line with full or partial occlusions in them. In case of document image recognition, a set of printed and hand written documents are digitized through a scanner by manually placing the document on the bed of the scanner.

### 2.4 Image preprocessing

The images for all the domains of application suffer from slight rotation due to the process of capturing itself. A rotation algorithm is applied to make them horizontally aligned. The color images are then converted to grey level images by finding the grey value of each pixel loated at *(i,j)* from the 24-bit color value of it using the following formula:

$$grey(i, j) = 0.59 \bullet red(i, j) + 0.30 \bullet green(i, j) + 0.11 \bullet blue(i, j)$$

### 2.5 Image binarization

The preprocessed grey level image is banarized using any binarization algorithm. Keeping in mind the variety of application domains Otsu's algorithm is applied here for the purpose of binarization. The binarized image is then stored as a bmp file for visual interpretation.

### 2.6 Edge detection

The edge map of the grey image is generated using Sobel's edge operator [5]. Mathematically, the operator uses four 3×3 kernels which are convolved with the original image to calculate approximations of the derivatives - one for horizontal direction, one for vertical direction and two for the two diagonal directions. The masks used are given below.

$$\begin{bmatrix} +1 & +2 & +1 \\ 0 & 0 & 0 \\ -1 & -2 & -1 \end{bmatrix}, \begin{bmatrix} -1 & 0 & +1 \\ -2 & 0 & +2 \\ -1 & 0 & +1 \end{bmatrix} \begin{bmatrix} 0 & +1 & +2 \\ -1 & 0 & +1 \\ -2 & -1 & 0 \end{bmatrix} \begin{bmatrix} -2 & -1 & 0 \\ -1 & 0 & +1 \\ 0 & +1 & +2 \end{bmatrix}$$

All four masks (horizontal vertical and two diagonal) are used for the detection of edge gradient. The edge gradient is then binarised using any binarization algorithm to geneate the binarized edge map of the image.

### 2.7 Hough transform for line extraction

Hough transform is applied on the binarized edge map to generate the Hough image of it. For this purpose, the parameters of the Hough transform, like deltaRo, deltaTheta, startTheta, endTheta, connectDistance and pixelsCount are initialized or tuned in such a way that the lines are extracted as a set of connected words. In the current work, any pixel-level line having a skew angle of $85^0$ to $95^0$ is considered with deltaTheta taken as $1^0$. The connectDistance and pixelsCount values are kept as 50 and 30 respectively. The Hough image is stored as bmp file for analysis of the performance of the system.

### 2.8 CCL Algorithm

All the marked white band of lines in tho Hough image are segmented through CCL algorithm. In this algorithm, any white pixel searches for its white neighbours. The neighbours will also search for their white neighbours in a recursive way. In the present work, 4-connected neighbours are searched and non-recursive fundtion call is used to reduce usage of system resource and time complexity. Each and every connected component is labeled accordingly for future use in word segmentation section.

### 2.9 Hough transform for word extraction

Hough transform is applied on the line segments to generate the Hough image of it at the words level. For this purpose, the parameters of the Hough transform, like deltaRo, deltaTheta, startTheta, endTheta, connectDistance and pixelsCount are initialized or tuned in such a way that the words are extracted as a set of connected characters. In the proposed work, any pixel-level line having a skew angle of $30^0$ to $120^0$ is considered with deltaTheta taken as $1^0$. The connectDistance and pixelsCount values are kept as 20 and 2 respectively. The new Hough image is stored as bmp file for analysis of the performance of the system.

The CCL algorithm as described in section 2.7 is again applied on the Hough image to segment it terms of words.

### 2.10 Bounding box creation

The location of the word segments as generated during the time of running CCL algorithm are stored in terms of starting row, starting column, ending row and ending column of the component. The locations of the segments are marked in terms of bounding box drawn over the binary image. The boxed binary image is then stored as a bmp file as an outcome of the system as well as for evaluation of the performance of it.

## 3 EXPERIMENTAL PROTOCOL

The present work is carried over a variety of standard dataset. The web uploaded handwritten scanned document image dataset for Bengali and English scripts are downloaded from www.cmaterju.org, the official website of CMATER, JU. These along with some other document images constitute the dataset for word and line segmentation from document images. Some type written documents are also printed and the scanned to incorporate variety in the document image dataset. The dataset contains colored texts also with or without background shades. The proposed segmentation algorithm is applied over the dataset to segment the words therein such that an OCR system can generate an editable text database from the non-editable document images.

The bisuness cards of persons from academics and industries are collected. The photo snaps of these cards are taken using hand held mobile phones. These images are then transferred to the computer through bluetooth device to generate the dataset for BCR. The current segmentation algorithm is applied over the dataset to segment the words therein such that an OCR system can generate an editable record of person from the non-editable document images.

The dataset for LPR system is collected from an ongoing project on red light violation detection system of a state government organization of India. There the images are captured at regular interval of time throughout the day and night using surveillance camera placed at a road crossing in a reputed metro city in India. The basic objective of the system is to localize and recognize the license plate of the violating vehicles. The proposed algorithm plays a major role in localizing the license plate containing the license number as a sequence of texts. Though the basic object of the proposed work is to identify the texts within an image, the efficiency of the algorithm can be enhanced just by tuning the Hough parameters for different domain of applications. The tuning parameters have already been discussed in sections 2.6 and section 2.8.

## 4 RESULTS AND DISCUSSIONS

### 4.1 Document image segmentation

For segmentation of text from document images, 15 Bengal script pages, 15 English script pages and 15 mixed script pages are used. Total 45 documents contain 812 lines consisting of 7308 words. Fig. 3. shows the results of text segmentation from document image by the proposed method. Each row represents the processing steps of a particular document image whereas each column indicates the output of each processing step. The figures in the first column show the binarized image, the figures in the second column show the line extraction in the Hough image space, the figures in the third column show the word extraction in the Hough image space and the figures in the fourth column show the word segments outbounded by black boxes in the binarized image space.

### 4.1 BCR image segmentation

For text segmentation from business cards to be used for a BCR system mobile camera captured images are used. In this proposed work 20 business cards are collected and digitized through a mobile camera. The cards consist of 165 lines as a whole with 990 words within them. Usually the words within the cards are a combination of texts with different font face, font size and font type. They may have different foreground colors and background shades also. The cards may also contain logo, image and other types of graphics also along with the text. The proposed method segments the proper text regions from the mixture of text and graphics with variation in the text pattern itself. Fig. 4. shows the results of text segmentation from BCR images by the proposed method. Each row represents the processing steps of a particular BCR image whereas each column indicates the output of each processing step. The figures in the first column show the binarized image, the figures in the second column show the word extraction in the Hough image space and the figures in the third column show the word segments outbounded by black boxes in the binarized image space.

### 4.1 LPR image segmentation

The segmentation of text from LPR images is a very difficult task as the images are captured autonomously by the camera irrespective of the natural scene in front of it. The natural scene at a road crossing may consists of a number of vehicles, general public, road side hoardings etc. As the primary objective of an LPR system is to localize the license plate and recognise the license number, the objective of the proposed segmentation process is restricted to find the word segments only that consist of the license plate or the license number thereby eliminating other texts usually written in the body of the vehicle or in the hordings at road side. In the proposed work 50 LPR images are considered consisting of 76 license plate words. Fig. 5. shows the result of text segmentation from LPR images using the proposed method. Each row represents the processing steps of a particular LPR image whereas each column indicates the output of each processing step. The figures in the first column show the grey image, the figures in the second column show the word extraction in the Hough image space and the figures in the third column show the word segments outbounded by black boxes in the binarized image space. As seen from Fig. 5. that the Hough image contains some extra segments generated due to some texts in the image along with the segment for license plate. Considering the preimposed condition of finding the license number location within the image these extra segments are eliminated from consideration by using some characteristics of the license plate. If the condition of finding the location of license plate is not imposed then the system generates all the text segments within the image.

□

## 5 CONCLUSION

Though Hough transform has already been used by the research community for extracting the standard shapes within an image, its immense potential in line and word segmentations are utilised in the current work. The computational complexity is reduced by properly choosing the direction of segmentation in the document image simply by tuning the Hough parameters. When applied over a large variety of document image dataset, during the process of line segmentation the current work has successfully segmented 88% lines, over segmented 10% lines and under segmented 2% lines. Though the rate of under segmentation is low but there is considerable amount of over segmentation of line mainly because of the large and non-uniform separation of some of the words in handwritten document images. During word segmentation process the current method successfully segmented 85.7% words, over segmented 12.1% words and under segmented 2.2% words. Here also the non-uniformity of the inter-character spacing in a word makes some of the words over segmented in case of handwritten document images. It is observed that in some of the cases of handwritten document images the inter-character spacing is than the inter-word spacing. The algorithm fails to segment properly in case of very closely spaced lines. Some technique must be used as a post processing step to isolate touching text lines in the Hough image. The binarization may also play a crutial role in the process of segmentation. A good binarization technique may eliminate some of the document image segmentation problems.

In case of BCR images more emphasis is given in word segmentation rather than line segmentation. The current technique has successfully segmented 94.6% words, over segmented 4.4% words and under segmented 1% words. The result appears to be very encouraging in automatic processing of business cards for database entry as it is providing higher accuracy even if there is huge variation of printed texts within the business cards.

In case of LPR images according to the preimposed condition the intention was to segment the license plates of the vehicles from the surveillance camera image, the efficiency is calculated in terms of finding the license plate in the image as a text segment. The current technique successfully localizes 85.5% true license plates only. In 10% cases it localizes other text regions along with the license plate characters and in 4.5% cases it localizes texts other than the license plate characters. If the preimposed condition of finding the license plate only from the image is removed and the objective is to find any text within an image then the efficiency of the proposed technique reaches 88%. As the proposed technique is mainly for segmentation of texts from the images, this result appears to be quite satisfactory as the general text segments appearing along with the license plate segment can be easily removed from further consideration by incorporating a post processing module in the LPR system.

All the aforementioned results show that the proposed technique can be efficiently utilized in case of various domains of image segmentation which will be subsequently used in various domain specific OCR systems.

|   |   |   |   |
|---|---|---|---|
| (a) | (b) | (c) | (d) |
| (e) | (f) | (g) | (h) |
| (i) | (j) | (k) | (l) |
| (m) | (n) | (o) | (p) |
| (q) | (r) | (s) | (t) |

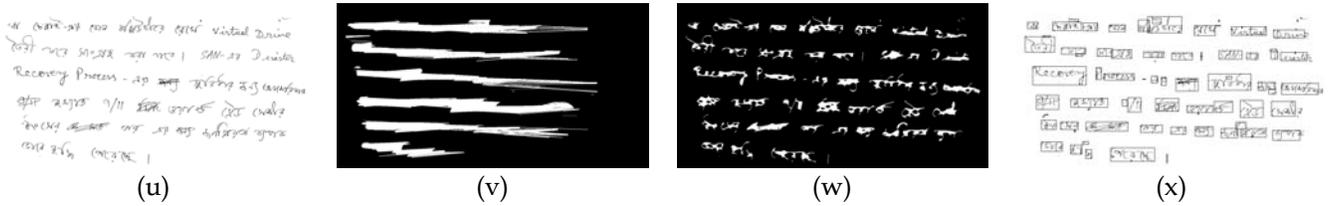

(u) (v) (w) (x)

Fig. 3. Text segmentation from multi-script handwritten document images

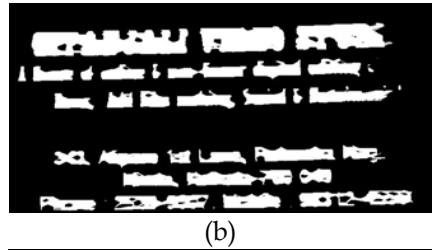

(a) (b) (c)

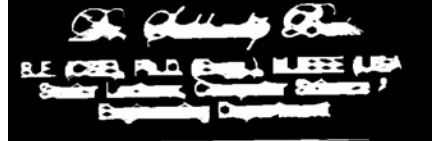

(d) (e) (f)

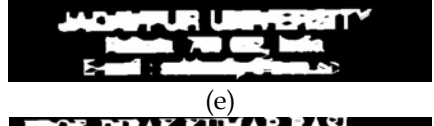

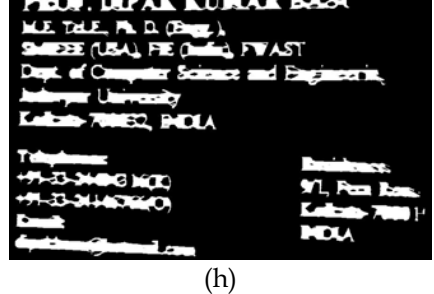

(g) (h) (i)

Fig. 4. Text segmentation from business card images

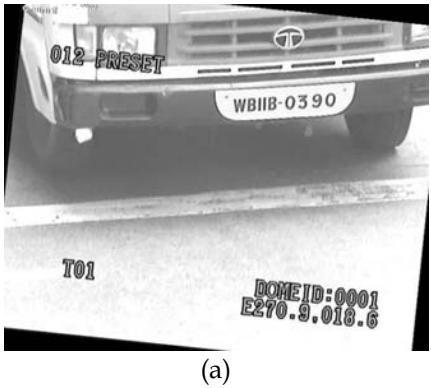 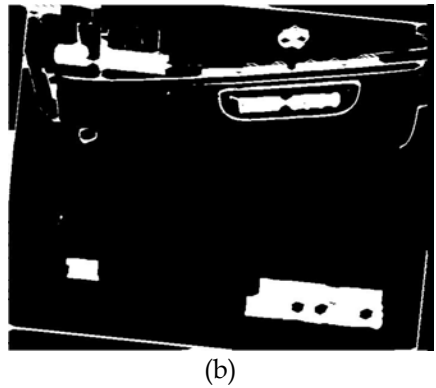 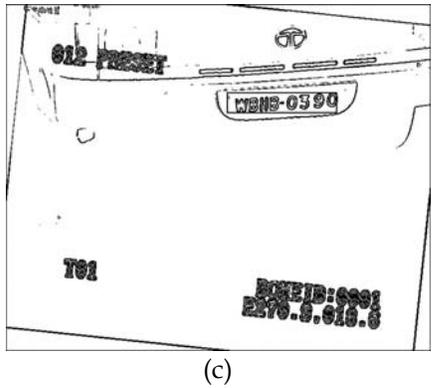

(a) (b) (c)

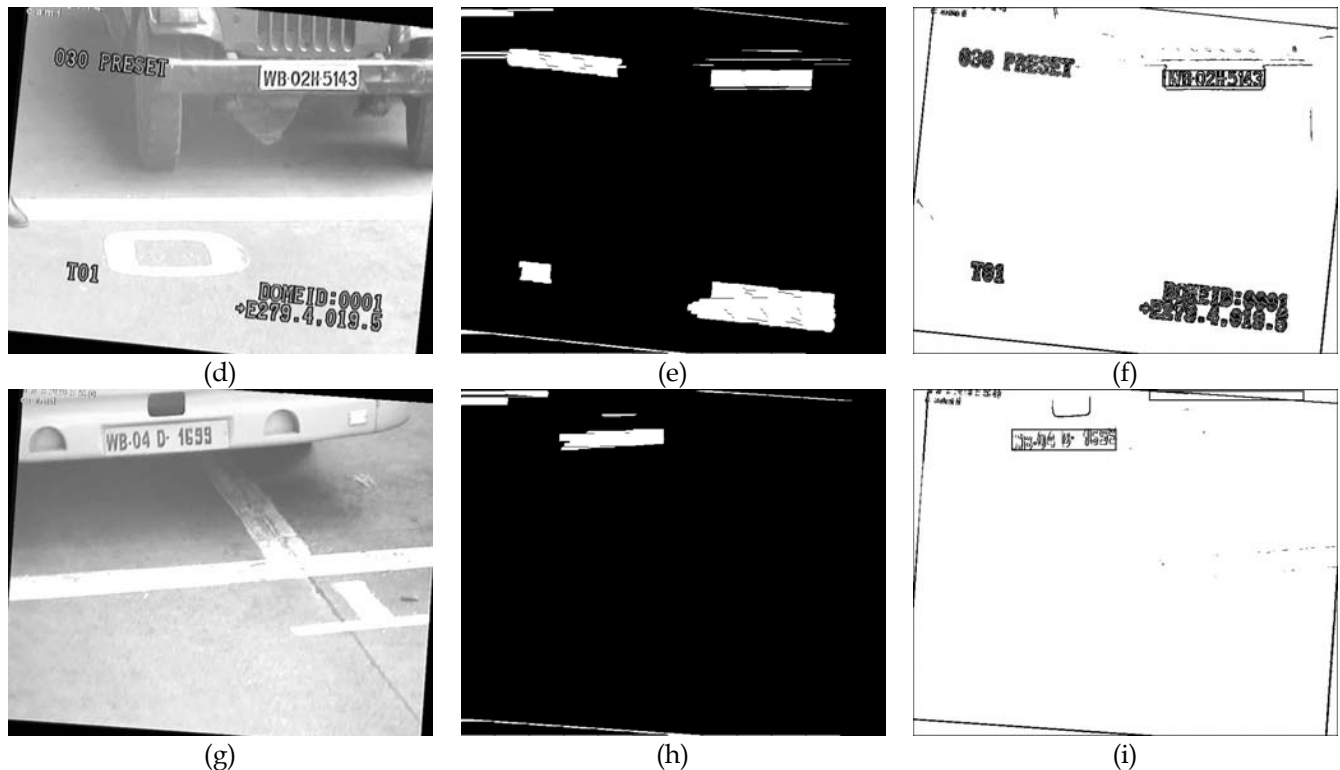

Fig. 5. Text segmentation from vehicle images obtained from an LPR systm

## ACKNOWLEDGMENT

Authors are thankful to the CMATER and the SRUVM project, C.S.E. Department, Jadavpur University, for providing necessary infrastructural facilities during the progress of the work. One of the authors, Mr. S. Saha, is thankful to the authorities of MCKV Institute of Engineering for kindly permitting him to carry on the research work.

## REFERENCES

[1] Anoop M. Namboodiri, Anil K. Jain, "Online Handwritten Script Recognition", *IEEE Trans. on Pattern Analysis and Machine Intelligence*, vol. 26, no. 1, pp. 124-130, Jan 2004.

[2] S. Basu, C. Chaudhuri, M. Kundu, M. Nasipuri, D.K. Basu, "Text line extraction from multi-skewed handwritten documents", *Pattern Recognition*, vol. 40, no. 6, pp. 1825-1839, Jun. 2007.

[3] A. Khandelwal et. al., "Text Line Segmentation for Unconstrained Handwritten Document Images Using Neighborhood Connected Component Analysis", S. Chaudhury et al., eds., PReMI 2009, Berlin: Springer-Verlag, pp. 369-374, 2009

[4] Satadal Saha, Subhadip Basu, Mita Nasipuri and Dipak Kumar Basu, "License Plate localization from vehicle images: An edge based multi-stage approach", *International Journal on Recent Trends in Engineering (Computer Science)*, vol. 1, no. 1, 2009, pp. 284-288.

[5] R. C. Gonzalez and R. E. Woods, *Digital Image Processing*, Second Edition, Pearson Education Asia, 2002.

**Satadal Saha** has done B. Sc. (Physics Hon's), completed B. Tech. in Applied Physics and M. Tech. in Optics and Optoelectronics from University of Calcutta in 1995, 1998 and 2000 respectively. In 2000, he joined as a Project Fellow in the Department of CST in BESU (formerly B. E. College), Howrah. In 2001, he joined as a Lecturer in the Department of Information Technology, Govt. College of Engg. and Textile Technology (formerly known as College of Textile Technology), Serampore. In 2004, he joined as a Lecturer in the Department of Computer Science and Engg., MCKV Institute of Engg, Howrah and he is continuing his service there as a Sr. Lecturer. He has authored a book titled Computer Network published by Dhanpat Rai and Co. Ltd., New Delhi in 2008. His research areas of interest are image processing and pattern recognition. Mr. Saha is a member of IEEE, IETE and CSI. He was also a member of the executive committee of IETE, Kolkata for the session 2006-08.

**Subhadip Basu** received his B.E. degree in Computer Science and Engineering from Kuvempu University, Karnataka, India, in 1999. He received his Ph.D. (Engg.) degree thereafter from Jadavpur University (J.U.) in 2006. He joined J.U. as a senior lecturer in 2006. His areas of current research interest are OCR of handwritten text, gesture recognition, real-time image processing.

**Mita Nasipuri** received her B.E.Tel.E., M.E.Tel.E., and Ph.D. (Engg.) degrees from Jadavpur University, in 1979, 1981 and 1990, respectively. She has been a faculty member of J.U since 1987. Her current research interest includes image processing, pattern recognition, and multimedia systems. She is a senior member of the IEEE, U.S.A., Fellow of I.E (India) and W.B.A.S.T, Kolkata, India.

**Dipak Kumar Basu** received his B.E.Tel.E., M.E.Tel., and Ph.D. (Engg.) degrees from Jadavpur University, in 1964, 1966 and 1969 respectively. He has been a faculty member of J.U from 1968 to January 2008. He is presently an A.I.C.T.E. Emiretus Fellow at the CSE Department of J.U. His current fields of research interest include pattern recognition, image processing, and multimedia systems. He is a senior member of the IEEE, U.S.A., Fellow of I.E. (India) and W.B.A.S.T., Kolkata, India and a former Fellow, Alexander von Humboldt Foundation, Germany.